# AntiPlag: Plagiarism Detection on Electronic Submissions of Text Based Assignments

MAC Jiffriya[1], MAC Akmal Jahan[1], Roshan G Ragel[2] and Sampath Deegalla[2]
[1]Post Graduate Institute of Science, University of Peradeniya
[2]Faculty of Engineering, University of Peradeniya

*Abstract* - **Plagiarism is one of the growing issues in academia and is always a concern in Universities and other academic institutions. The situation is becoming even worse with the availability of ample resources on the web. This paper focuses on creating an effective and fast tool for plagiarism detection for text based electronic assignments. Our plagiarism detection tool named AntiPlag is developed using the tri-gram sequence matching technique. Three sets of text based assignments were tested by AntiPlag and the results were compared against an existing commercial plagiarism detection tool. AntiPlag showed better results in terms of false positives compared to the commercial tool due to the pre-processing steps performed in AntiPlag. In addition, to improve the detection latency, AntiPlag applies a data clustering technique making it four times faster than the commercial tool considered. AntiPlag could be used to isolate plagiarized text based assignments from non-plagiarised assignments easily. Therefore, we present AntiPlag, a fast and effective tool for plagiarism detection on text based electronic assignments.**

*Index Terms*-Plagiarism Detection, AntiPlag, Electronic Assignments.

## I. INTRODUCTION

Plagiarism is known as illegal use of others' part of work or whole work as one's own. Plagiarism diminishes one's innovative thinking, creativeness and improvement of knowledge and also it is considered as illegal act in a moral society. In a survey that was conducted on plagiarism in the academia by the University of California in Berkley, it was shown that the percentage of plagiarism has increased by 74.4% within four years period (1993 – 1997) [1] and in another study by Butakov and Scherbinin concludes that more than 90.0% of high school students are involved in plagiarism [2].

Plagiarism is one of the growing issues in academia. Academic staff faces difficulties in marking students' assignments with higher degree of judgment and waste their valuable time for plagiarism detection. The paper focuses on building an effective, simple and fast tool for plagiarism detection on text based electronic assignments to minimize this issue and to help the academic staff in conducting proper evaluation of assignments.

The rest of the paper is organized as follows: In Section II of the paper, we categorise plagiarism detection into two and summarise the existing tools under each category. In addition, we identify one of the best plagiarism detection techniques for text based plagiarism detection and the major drawback of the current tools and therefore our contribution in this paper. In Section III, we present the methodology adapted to develop AntiPlag and in Section IV we present the results obtained from AntiPlag on some test plagiarism detection data. In Section V, we conclude the paper.

## II. RELATED WORK

Plagiarism can be classified as source code based plagiarism and free text based plagiarism. Early days most of the researchers focused on source code plagiarism and several tools were developed for detecting source code based plagiarism. Some such tools are Plagio Guard [3][4], JPlag [5], Moss [6], Saxon [9], Detecta Copius [7], Sherlock [8], Copy/Paste Detector (CPD) and Big Brother [9]. Recently there are several text based plagiarism detection tools that support detecting only intra-corpal plagiarism such as Dupli Checker [23] and Article Checker [24] or detecting only extra-corpal plagiarism such as Anti-P [10] or both such as Plagiarism Checker X [25], Turnitin [22] and Ferret [9]. Corpus here is the written document subjected to the particular plagiarism detection. In extra-corpal plagiarism detection, the test material is compared with material from outside such as the web resources, whereas intra-corpal plagiarism detection is performed by comparing materials within the learning community, thus original and plagiarized materials are found at the same place [11].

Some plagiarism detection tools provide web based services and some serve as standalone applications. Turnitin, Article Checker and Dupli Checker are examples for web based plagiarism detection services. These detection tools except Turnitin provide free and online text based plagiarism detection service and they check the test material with the sources available on the web. Even though Turnitin supports both extra-corpal and intra-corpal plagiarism detection it does not provide a free service [12][13]. Similarly Plagiarism Detector, Plagiarism Checker X, CopyCatch [9], WORDCheck [9], CopyFind [9] and Ferret [14][15] are standalone application software for text based plagiarism detection. Plagiarism Detector and Plagiarism Checker X are commercially available in the Internet.

Plagiarism detection techniques are varying in approach. Most of the researchers use N-gram technique as a base and improve it in variety of text based applications [16][17]. In information retrieval, precision and recall make much sense in calculating accuracy [11]. Through the experiments to define the best N-gram level for plagiarism detection, tri-gram and bi-gram show better results than other N-grams. Tri-gram based search is more rigid and has shown better precision while bi-gram based search is more flexible and shows better recall. Therefore, tri-gram sequence matching is an effective technique to develop a plagiarism detection tool due to the good value of precision [18]. Therefore, we have selected the same technique for AntiPlag with additional features to make AntiPlag both effective and fast.

One of the major problems with the current tools is that the time they take to perform the plagiarism detection. They consume large amount of computing time and therefore the teachers do not like to use them. Our objective here is to build a tool that is simple, effective and fast. Later in this paper, we will show how our tool outperforms a commercially available tool for plagiarism detection in latency while maintaining the same level of detection accuracy. Therefore, our contribution in this paper is a simple, fast and effective plagiarism detector that outperforms a commercially available tool for text based plagiarism detection.

### III. METHODOLOGY

AntiPlag is developed using the tri-gram sequence matching technique with the help of a scripting language. Figure 1 depicts the steps followed in the development and such steps are discussed in details under the subsections in this section.

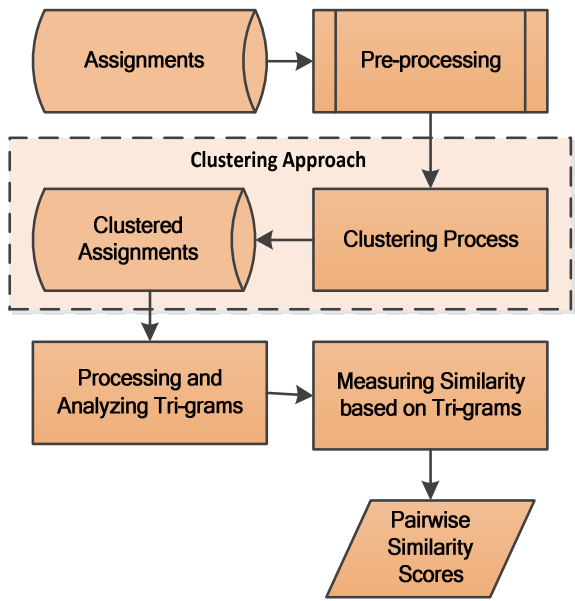

Fig.1: The two phase process of AntiPlag: Clustering and Pairwise Similarities

The electronic assignments were pre-processed before they are sent through a clustering algorithm. We use clustering here as an approach to expedite the plagiarism detection process and claim as one of the contributions. Therefore, we have developed two versions of AntiPlag, one that performs clustering and the other that does not. Later, we compared the effect of clustering on the detection latency by performing the tests with both the versions of AntiPlag. The step that follows clustering is the tri-gram construction and analysis. The tri-gram analysis is used for measuring pairwise similarities among the assignments tested and the results are presented to the user as percentage scores representing similarities.

#### A. Data Collection and File Conversion

Electronic text based assignments are collected as three isolated datasets. They were different in format and have been converted into plain text format to maintain all documents in the same format for fair treatment.

#### B. Pre-processing

The pre-processing step we performed, we consider as an important step for the plagiarism detection process. The purpose of pre-processing is to form suitable data which is to be input into the detection processes and to increase the effectiveness of the plagiarism detection tool. The corpus consists of a mixture of lower case and upper case characters. All upper case characters in the corpus have been transformed into lower case characters to eliminate case sensitivity from the corpus. All the diagrams, pictures and images in the corpus have been removed. The delimiters in the corpus and stop words have been identified and removed from the corpus.

#### C. Tri-gram Construction

A tri-gram consists of three consecutive words sequence in each line. The tri-gram sequences are formed from the pre-processed assignments. Tri-grams have been constructed by extracting collection of tri-gram sequences from the corpus. Figure 2 depicts the formation of tri-grams for an example sentence.

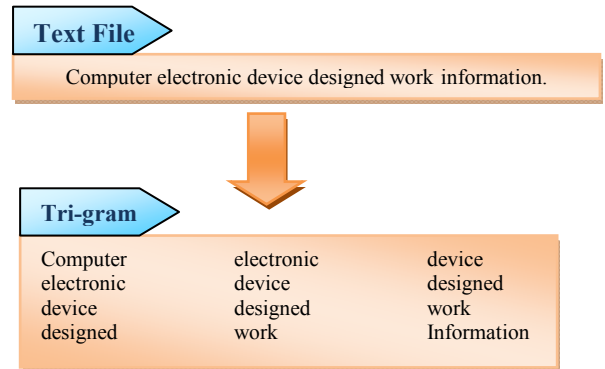

Fig. 2: Construction of Tri-gram

#### D. Similarity Measure

Tri-gram sequences of each pair of the assignment are compared using a tri-gram sequence matching technique and the similarity has been measured in percentage. Therefore, the measurement is that the higher the percentage scores the higher the similarity.

#### E. Clustering

To improve the detection latency of AntiPlag, a clustering approach was applied on the data sets to create appropriate clusters using K-Means algorithm using the WEKA tool. WEKA is an open source tool with collection of data mining algorithms [21].

In the clustering approach, the datasets have been passed into pre-processing. String to word vector was used to convert string attributes into a set of bag of words. K-Means algorithm was applied to the dataset since it has several advantages over document clustering [19][20]. Min-term frequency is changed and the dataset is evaluated using K-Means algorithm with varying cluster numbers K. The suitable cluster number K was selected from the experimental results for the datasets based on percentage of incorrectly clustered instances of the assignments.

*F. Stemming*

A stemming technique is applied to the datasets to convert the bag of words into their root words to test how much stemming process affects the plagiarism performance. Porter-Stemming algorithm [26] is used to stem all. Stemmed data sets are tested on AntiPlag tool again and the both results are compared.

## IV. RESULTS AND DISCUSSION

Three data sets of electronic assignments were tested by AntiPlag and the results were compared with existing commercial plagiarism detection tool known as "Plagiarism Checker X".

Figures 3, 4 and 5 represent maximum percentage of plagiarism of each assignment in AntiPlag and Plagiarism Checker X for Data-Set1, Data-Set2 and Data-Set3 respectively. AntiPlag showed better results in proper detection than the Plagiarism Checker X for all three data-sets due to the pre-processing steps used in AntiPlag.

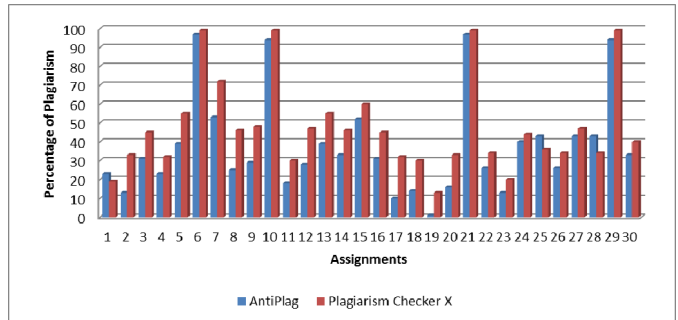

Fig. 3. Plagiarism Detected in AntiPlag vs. Plag. Checker X for Data-Set1

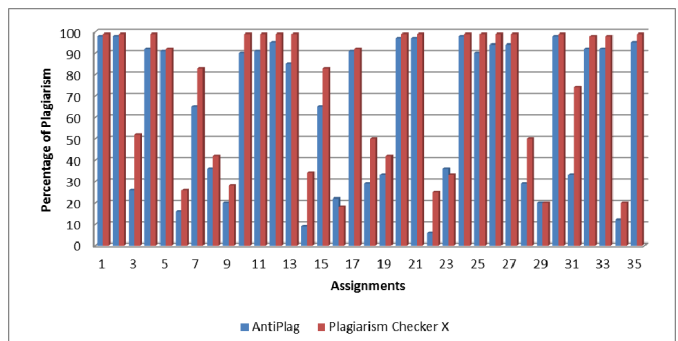

Fig. 4. Plagiarism Detected in AntiPlag vs. Plag. Checker X for Data-Set2

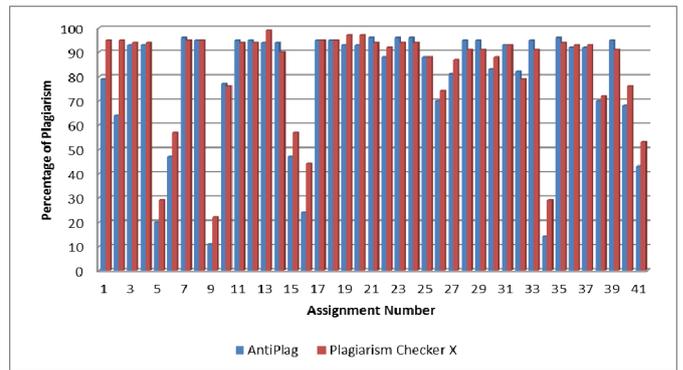

Fig. 5. Plagiarism Detected Rate in AntiPlag vs. Plag. Checker X for Data-Set3

All assignments in the three data-sets are stemmed and the datasets were tested on AntiPlag tool. The results of stemming datasets are compared with results of original data sets obtained without stemming. Figures 6, 7 and 8 represent percentage of plagiarism detected in AntiPlag with stemming and without stemming of datasets 1, 2, and 3 respectively.

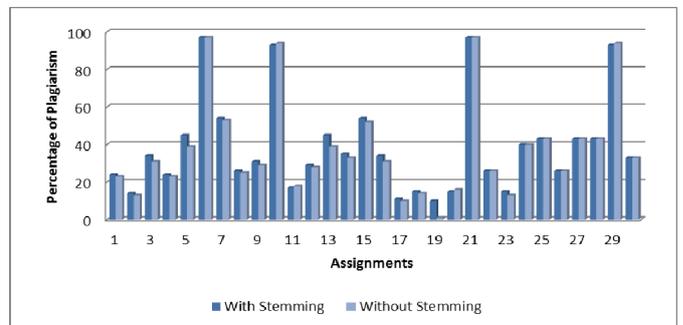

Fig. 6. Comparison of Plagiarism with and without Stemming of Data-Set1

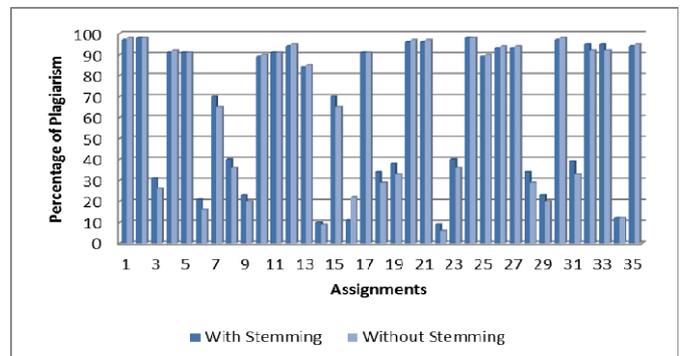

Fig. 7. Comparison of Plagiarism with and without Stemming of Data-Set2

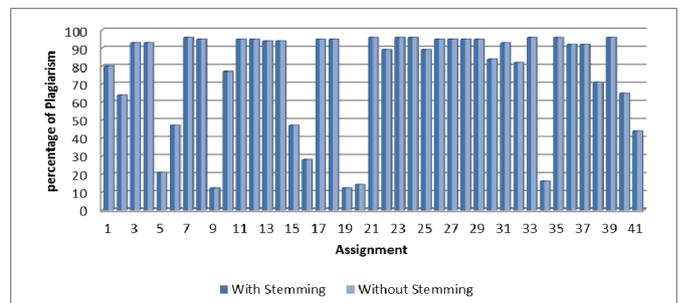

Fig. 8. Comparison of Plagiarism with and without stemming of Data-Set3

According to Figures 6, 7 and 8, there is no significant difference in percentage of plagiarism detected between datasets with and without stemming. Therefore, it was decided that the stemming process does not influence in finding plagiarism in text based assignments for the dataset considered.

As explained earlier, a clustering approach was used to improve the detection latency of plagiarism in text based assignments. The three datasets are clustered separately using K-Means clustering algorithm by using WEKA data-mining tool. Table 1 represents the clusters with the cluster numbers corresponding to the number of assignment datasets. Data-Set1 and Data-Set2 have been clustered into 25 clusters since they showed less percentage of incorrectly clustered instances in this point, whereas Data-Set3 showed less percentage of incorrectly clustered instances when it has been clustered into 20 clusters.

Table I: Cluster based Plagiarism of Three Data Sets

| Cluster No | Data-Set1 | Data-Set2 | Data-Set3 |
|---|---|---|---|
| 0 | 4 | 8 | 11, 12, 32 |
| 1 | 8 | 12, 35 | 21, 23, 24 |
| 2 | 10, 29 | 9 | 22, 25 |
| 3 | 28 | 34 | 17, 18, 33, 39 |
| 4 | 5, 13 | 5, 17 | 2, 28, 29 |
| 5 | 11 | 31 | 36, 37 |
| 6 | 14 | 14 | 13, 14 |
| 7 | 23 | 24 | 19, 20 |
| 8 | 2 | 23 | 41 |
| 9 | 30 | 15 | 3, 4 |
| 10 | 19 | 19 | 1, 10, 30 |
| 11 | 16 | 13 | 16 |
| 12 | 25 | 32, 33 | 9 |
| 13 | 12 | 16 | 5 |
| 14 | 1 | 22 | 15 |
| 15 | 17 | 7 | 34 |
| 16 | 15, 24 | 11 | 26 |
| 17 | 22 | 1, 30, 2 | 7, 8, 31, 35 |
| 18 | 3 | 26, 27, 4 | 27, 38, 40 |
| 19 | 20 | 29 | 6 |
| 20 | 6, 21 | 6 | |
| 21 | 7, 27 | 28 | |
| 22 | 18 | 3 | |
| 23 | 26 | 10, 20, 21, 25 | |
| 24 | 9 | 18 | |

According to the Table I, assignment numbers 5, 6, 7, 10, 13, 15, 21, 24, 27 and 29 are highly plagiarized in Data-Set1 while assignment numbers 1, 2, 4, 5, 10, 12, 20, 21, 25, 26, 27, 30, 32, 33 and 35 are highly plagiarized assignment in Data-Set2 compared to the rest of the assignments. In Data-Set3 the assignment numbers 5, 6, 9, 15, 16, 26, 34 and 41 are not plagiarized and others are highly plagiarized.

Figures 9, 10 and 11 depict percentages of plagiarism in some randomly selected pairs of assignments from Data-Set1, Data-Set2 and Data-Set3 respectively. As it can be observed, paired assignments within a cluster showed very high amount of plagiarism whereas paired assignments across clusters showed negligible amount of plagiarism in all three datasets. Therefore, we could conclude that the clustering approach is useful in separating plagiarized and non-plagiarized assignments easily.

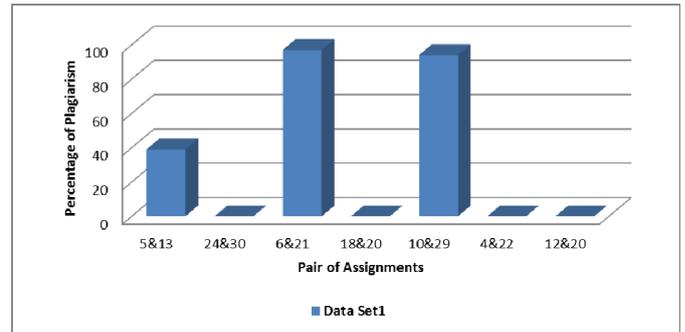

Fig. 9. Plagiarism detection based on clustering approach for Data-Set1

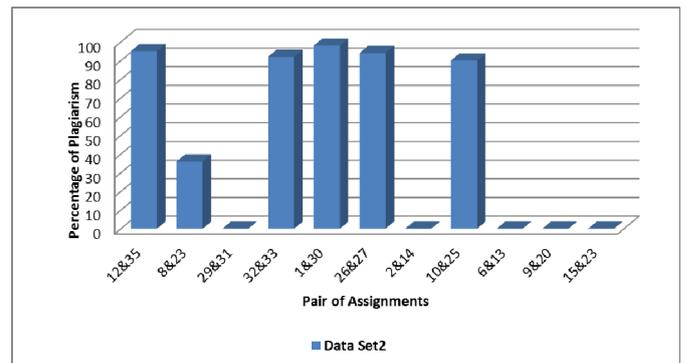

Fig. 10. Plagiarism detection based on clustering approach for Data-Set2

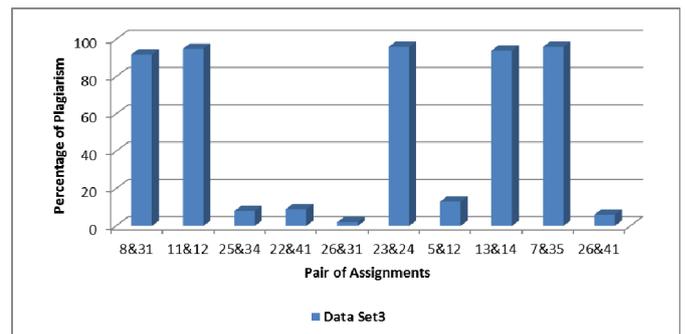

Fig. 11. Plagiarism detection based on clustering approach for Data-Set3

Figure 12 shows the execution time for plagiarism detection by the Plagiarism Checker X and the AntiPlag for the three datasets with and without the clustering applied. AntiPlag tool has taken lesser time for all three datasets than Plagiarism Checker X. According to the results in Figure 12, AntiPlag was more than two times faster than Plagiarism

Checker X before the clustering technique is applied. The clustering approach detailed earlier is applied to improve the performance (the detection latency) of the plagiarism detection. The same figure (Figure 12) shows the execution time of AntiPlag with clustering applied for all three datasets. It can be seen that the execution time of AntiPlag is improved by approximately two times when the clustering approach is applied for our data sets.

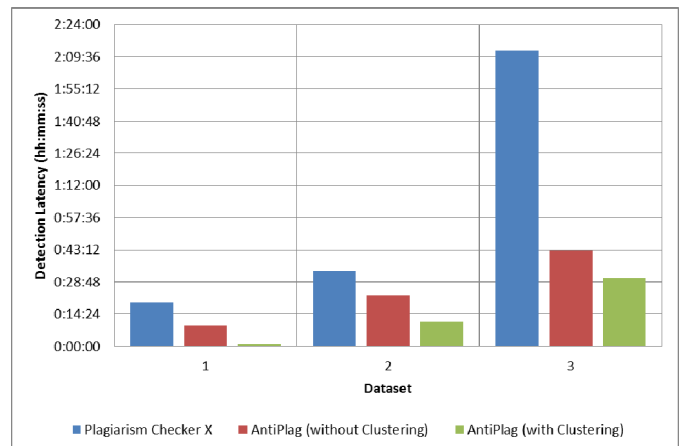

Fig. 12. Execution Time of Plagiarism Detection Tools for three Datasets

Therefore, it can be concluded that AntiPlag with clustering applied will have at least four times performance improvement (time taken to detect plagiarism) over the commercial tool named Plagiarism Checker X.

## V. CONCLUSIONS

We present a plagiarism detection tool named AntiPlag, where it is optimized and enhanced through the clustering approach. AntiPlag is fast and capable to compare all pair of assignments automatically at once. In addition, through experiments, we have proved that the cluster based AntiPlag is an effective, simple and fast tool for plagiarism detection on text based electronic assignment.